# Photoinduced magnetization change in multiferroic $YbFe_2O_4$

Chang-Hui Li,[1] Yi Liu,[1] Fen Wang,[1] Xuan Luo,[2] Yu-Ping Sun,[2] Xiang-Qun Zhang,[1] Zhao-Hua Cheng,[1] and Young Sun[1,a)]

[1]Beijing National Laboratory for Condensed Matter Physics, Institute of Physics, Chinese Academy of Sciences, Beijing 100190, P. R. China

[2]Key Laboratory of Materials Physics, Institute of Solid State Physics, Chinese Academy of Sciences, Hefei 230031, P. R. China

[a)]Electronic mail: youngsun@aphy.iphy.ac.cn

Abstract

We have studied the influence of laser illumination on the magnetization in multiferroic $YbFe_2O_4$ single crystals. A photoinduced magnetization change has been confirmed in both ab plane and c axis direction. The temperature dependence of the photoinduced magnetization reduction excludes laser heating as the cause. In terms of the breakdown of charge order driven by laser illumination, the photoinduced magnetization change provides a strong evidence for the spin-charge coupling in $YbFe_2O_4$. This photomagnetic effect based on charge-order-induced multiferroicity could be used for non-thermal optical control of magnetization.

Recently there has been growing interest in the field of multiferroics and magnetoelectric effects due to its great potential for applications as well as the significance for fundamental physics.[1,2] The mixed-valence compounds $RFe_2O_4$ (R= Ho, Er, Tm, Yb, Lu and Y) represent a peculiar class of multiferroic materials in which the multiferroicity is induced by charge order (CO).[3] The system involves a double layer structure with a triangular iron lattice. The coulombic interactions between $Fe^{2+}$ and $Fe^{3+}$ ions compete with the frustrated nature of the triangular lattice, which leads to a peculiar ordered arrangement of charges. For instance, a three-dimensional (3D) CO occurs below 330 K in $LuFe_2O_4$,[4,5] which results in a net electrical polarization. This type of ferroelectricity associated with CO is termed as "electronic ferroelectricity",[6] in contrast to conventional ferroelectricity involving displacement of cation and anion pairs. Meanwhile, the strong magnetic interactions between Fe moments develop as a ferrimagnetic ordering below a temperature around 240 K.[7] Therefore, the coexistence of electronic ferroelectricity and ferrimagnetism makes $RFe_2O_4$ a new class of multiferroic materials.

Although the CO induced multiferroicity in $RFe_2O_4$ has been well identified by many experiments, the coupling between electronic and magnetic degree of freedom in the system has not been directly observed. Previous studies have shown some signs of a magnetoelectric (ME) coupling in $LuFe_2O_4$, such as the sharp change of electrical polarization at the ferrimagnetic transition temperature $T_N$,[6] and the X-ray scattering study.[8] Nevertheless, the ME coupling in multiferroic $RFe_2O_4$ requires more convincing evidences.

It has been well known that the CO in many systems can be manipulated by external stimulus.[9,10] Our recent work suggests that the CO in $RFe_2O_4$ is very sensitive to external electric fields and the breakdown of CO by applied electric fields leads to an insulator to metal transition[11] as well as a dramatic change in magnetization.[12] Another way for manipulating the CO is by photoexcitation. As observed in a variety of perovskite manganites, the CO state can be melted by illumination with X-ray[13] or visible radiation.[14,15] A similar photoinduced melting of the CO state in $RFe_2O_4$ would be expected as its CO state is more sensitive to external stimulus. In this work, we have grown high quality single crystals of $YbFe_2O_4$, a prototype member of the $RFe_2O_4$ family, and studied the influence of laser illumination on the magnetization. The results confirm that the magnetization has apparent changes in response to the melting of CO driven by laser illumination, which indicates a strong ME coupling in $YbFe_2O_4$.

Single crystals of $YbFe_2O_4$ were grown by optical floating-zone melting method in a flowing argon atmosphere. X–ray diffraction at room temperature showed that the samples are single phase and have a structure consistent with literature. Thin slices of $YbFe_2O_4$ crystals with a size of 0.9 mm×0.9 mm×0.3 mm were used for the study. The magnetization was measured with a Quantum Design Superconducting Quantum Interference Device (SQUID) magnetometer using a optical-fiber sample holder. The source of illumination used in our experiments was a continuous work (cw) Nd-YAG laser with the wavelength $\lambda$=532 nm, photon energy $h\nu$=2.33 eV and the maximum power of 3 mW.

Figure 1 shows the temperature dependence of magnetization of $YbFe_2O_4$ in a 0.1 T field with both zero-field cooling (ZFC) and field-cooling (FC) processes. Due to the layered structure, a strong magnetic anisotropy is observed. The magnetization along c axis is much higher than that in the ab plane, suggesting that the easy axis is along c direction. For both directions, the magnetization increases fast around 250 K and shows a peak in the ZFC curve, which indicates a paramagnetic to ferrimagnetic transition.

To study the influence of laser illumination on magnetization, we first performed a measurement at 240 K. As shown in Fig. 2(a), the sample was cooled in zero field from 300 to 240 K, then a 0.1 T magnetic field was applied along c axis and the magnetization was measured as a function of time. After a time, a laser illumination is shed on the sample along c axis and the magnetization was measured simultaneously. Once a single data point was recorded, the illumination was immediately blocked and the magnetization was measured for another period. This process was repeated for several times for confirmation. It is clearly seen that the laser illumination does induce an apparent change in magnetization. The magnetization drops when the illumination is on, and returns when the illumination is off. One may argue that the photoinduced magnetization change could be due to laser heating. As seen in the *M-T* curve in Fig. 1, at 240 K, a rise in sample temperature can also cause a decrease in magnetization. In fact, when the laser illumination was kept on for a few minutes, the magnetization decreases slowly as seen in Fig. 2(a). This magnetization change with long-time laser illumination might be due to a heating effect. However, the small value of magnetization change implies that the rise in sample temperature is not significant even with continuous laser heating for a long time. Since the time window for collecting a single data point in the SQUID is about a few seconds, the temperature change by laser illumination in this short period should be quite small. Therefore, the apparent magnetization change is unlikely due to laser heating.

In order to clarify the origin of this photoinduced magnetization change, we performed another measurement at 220 K which is below the ZFC peak temperature. As shown in Fig. 2(b), the sample was cooled in zero field from 300 to 220 K, then a 0.1 T magnetic field was applied along c axis and the magnetization was measured as a function of time. Because the ZFC and FC magnetization has a big divergence at 220 K, there is a strong relaxation in magnetization. When the laser illumination is on, the magnetization drops, similar to that observed at 240 K. Since a rise of sample temperature at 220 K should cause an increase in magnetization (see the ZFC curve in Fig. 1), the magnetization drop induced by laser illumination can not be due to the heating effect. For comparison, we keep the laser illumination on for a few minutes and measure the magnetization simultaneously. In this case, after an initial drop, the magnetization does increase slowly with time, which might be due to laser heating effect. Therefore, the measurement at 220 K excludes laser heating as the origin of magnetization drop induced by laser illumination.

Considering the strong magnetic anisotropy, we also studied the influence of laser illumination on the ab-plane magnetization. In these measurements, both the magnetic field and laser beam are applied in the ab plane of the crystal. As shown in Fig. 3(a), the photoinduced magnetization reduction is also clearly seen at 240 K in a constant 0.1 T field. Since laser heating may also lead to a decrease of magnetization at this temperature, we performed another measurement at 200 K. As shown in Fig. 3(b), the magnetization drops when the laser illumination is on, which is opposite to the laser heating effect at this temperature.

All the above measurements were done in a constant magnetic field. We find that the photoinduced magnetization can also be observed in the remanent magnetization without a magnetic field. As shown in Fig. 4, the sample was cooled in a 5 T field from 300 to 200 K, then the field is removed and the remanent magnetization was measured with time. The magnetization drops when the laser illumination is on, and returns when the laser illumination is off. This is against what expected for laser heating. In zero magnetic field, the remanent magnetization should always decay down even with a temperature disturbance. If the magnetization drop is due to a temperature rise by laser heating, the magnetization should remain at the low state after the illumination is off. The increase of remanent magnetization after the laser illumination is off further proves that the photoinduced magnetization change is not due to laser heating.

As we mentioned in the introduction, the role of laser illumination is to manipulate the CO state. The breakdown of the CO by light illumination has been widely observed in many CO systems, which usually leads to an increase in conductivity and even an insulator to metal phase transition.[14,15] In addition, a photoinduced magnetization change has been observed in several charge ordered manganites.[16,17] We think that a photoinduced breakdown of CO also happens in $YbFe_2O_4$. In fact, the (partial) melting of CO by laser illumination in $RFe_2O_4$ can be supported by the photocurrent effect.[18] In terms of the melting of CO by laser illumination, the photoinduced magnetization change indicates a strong spin-charge coupling in multiferroic $YbFe_2O_4$. Once the CO is broken by laser illumination, the ferrimagnetic order is also destroyed due to a spin-charge coupling, which results in a decrease in magnetization. The physical process of the spin-charge coupling in $RFe_2O_4$ deserves further studies, though there has been some theoretical work discussing the possible mechanism.[19]

We note that though the photoinduced magnetization change is well detectable, the relative change of magnetization is small (~ 4%). This is due to the fact that we used bulk single crystal samples. As the thickness of the samples is 0.3 mm which is far greater than the penetration depth of the laser illumination, only a small proportion of the total magnetization at the surface could be affected. An amplified effect would be expected in thin film samples.

In summary, we have studied the influence of laser illumination on the magnetization of multiferroic $YbFe_2O_4$. A photoinduced magnetization reduction has been confirmed in the ferrimagnetic state. The temperature dependence of this effect excludes laser heating as the cause. Instead, this photoinduced magnetization change can be understood in terms of an intrinsic spin-charge coupling in response to the CO breakdown by laser illumination. Therefore, this intrinsic photomagnetic effect could be used for nonthermal optical control of magnetization.

This work was supported by the National Natural Science Foundation of China (50721001, 50831006) and the National Key Basic Research Program of China (2007CB925002).

Fig. 1 Temperature dependence of magnetization of $YbFe_2O_4$ in a 0.1 T field with ZFC and FC processes.

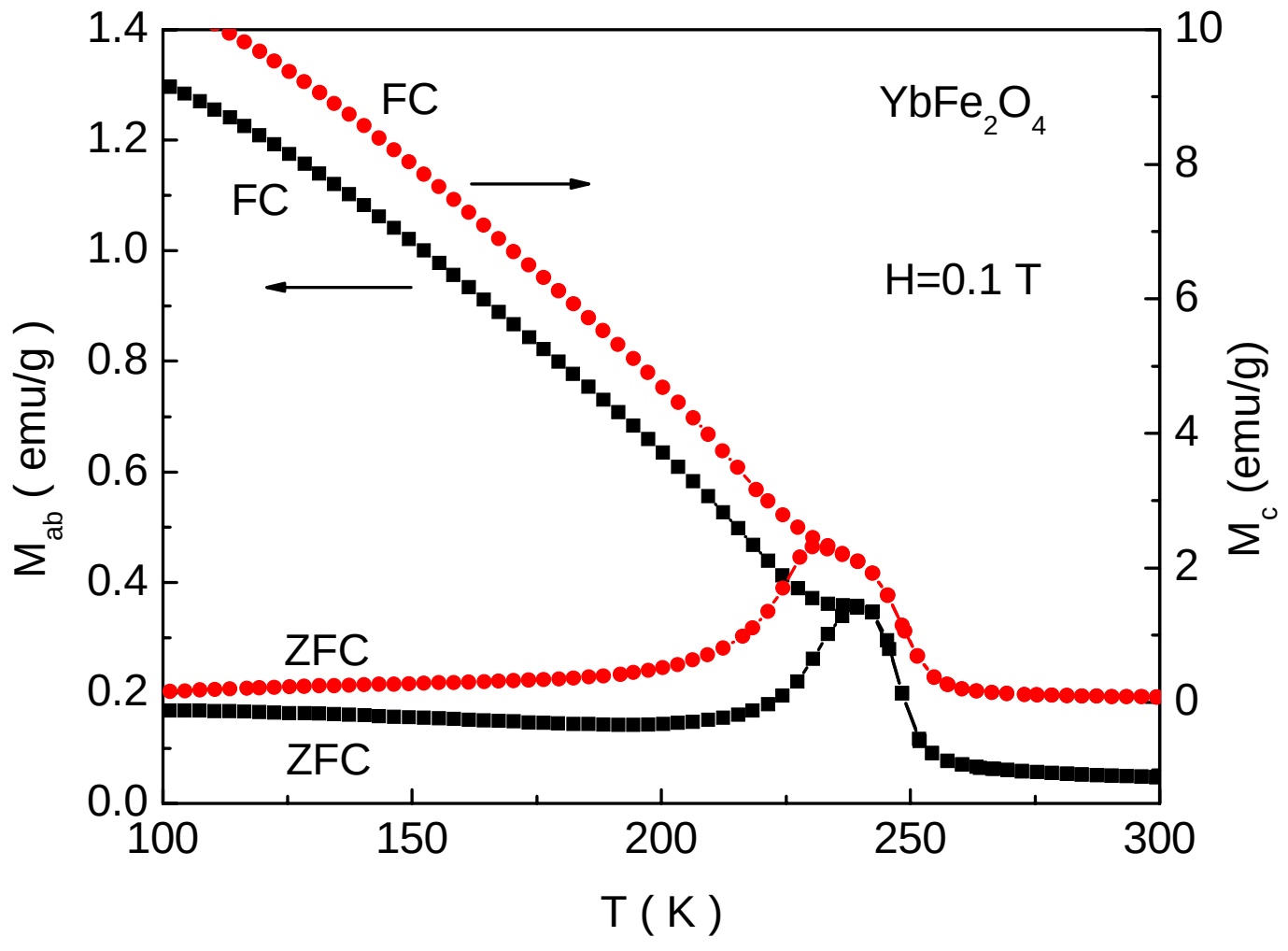

Fig. 2 The c-axis magnetization as a function of time in a 0.1 T field at (a) 240 K and (b) 220 K. The red bars mark the time windows during which the laser illumination is on. The inset shows the measurement configuration. The illumination and magnetic field are applied along c direction.

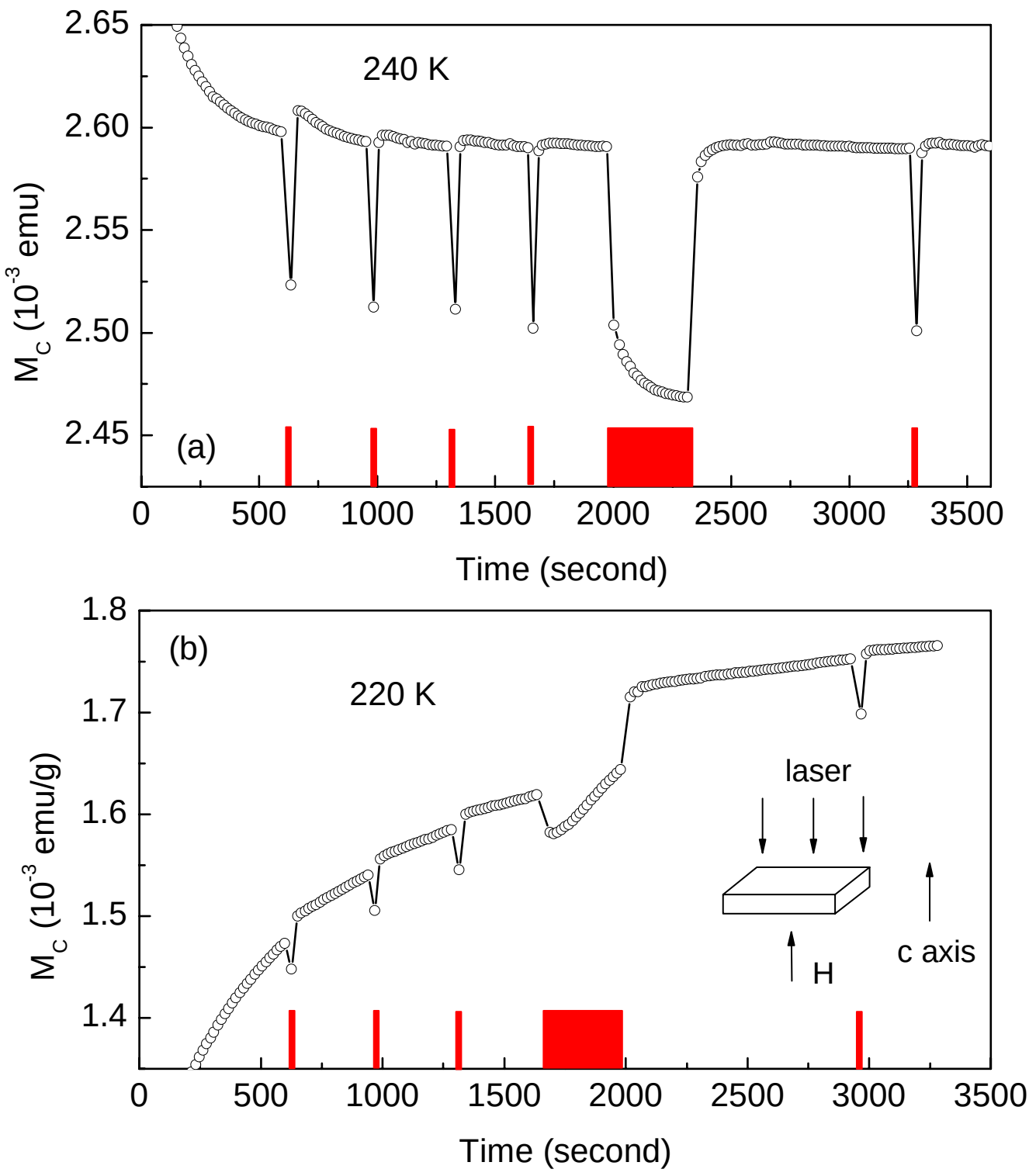

Fig. 3 The ab-plane magnetization as a function of time in a 0.1 T field at (a) 240 K and (b) 200 K. The red bars mark the time windows during which the laser illumination is on. The inset shows the measurement configuration. The illumination and magnetic field are applied in ab plane.

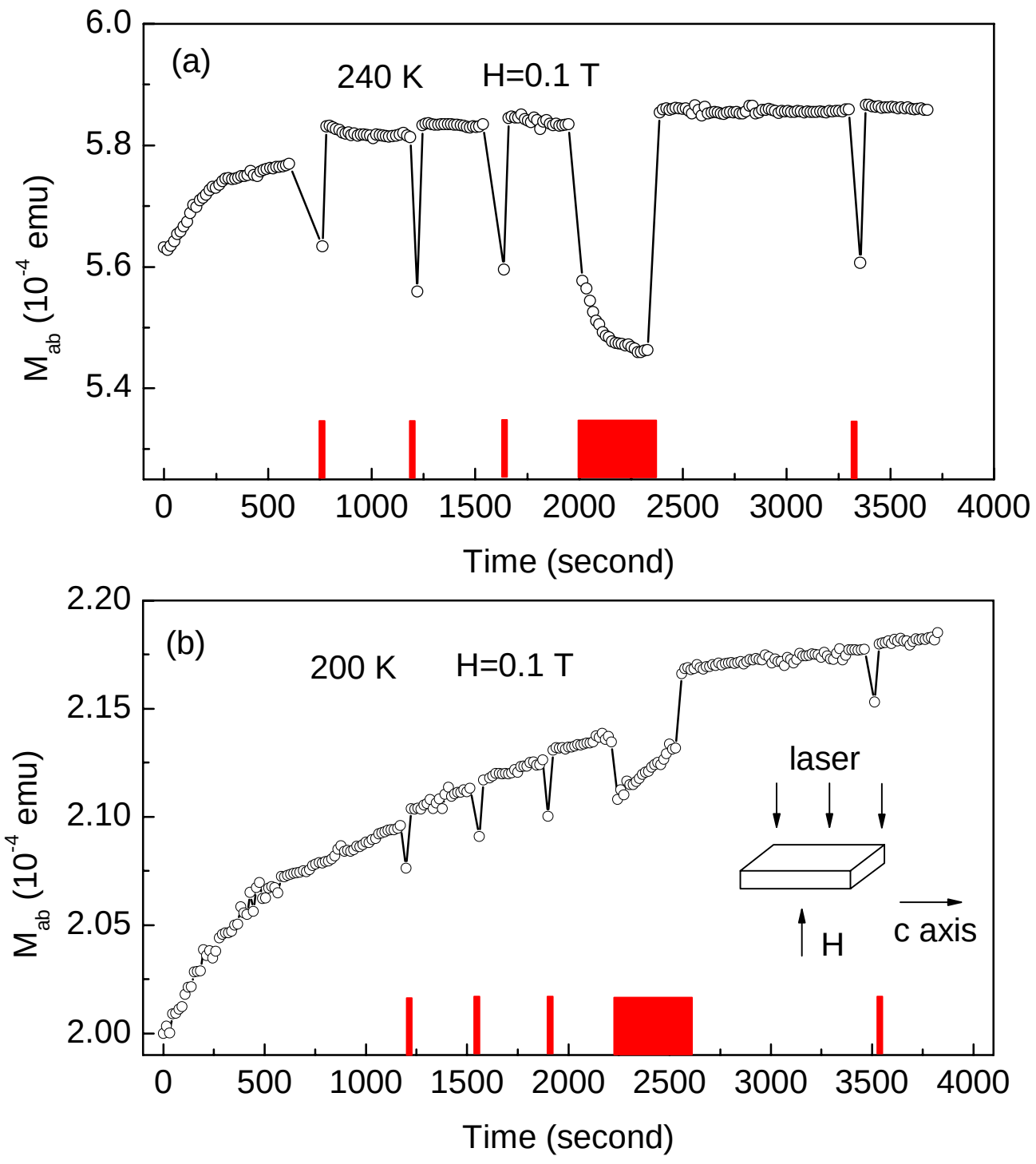

Fig. 4 The remanent magnetization of ab plane as a function of time at 200 K. The red bars mark the time windows during which the laser illumination is on.

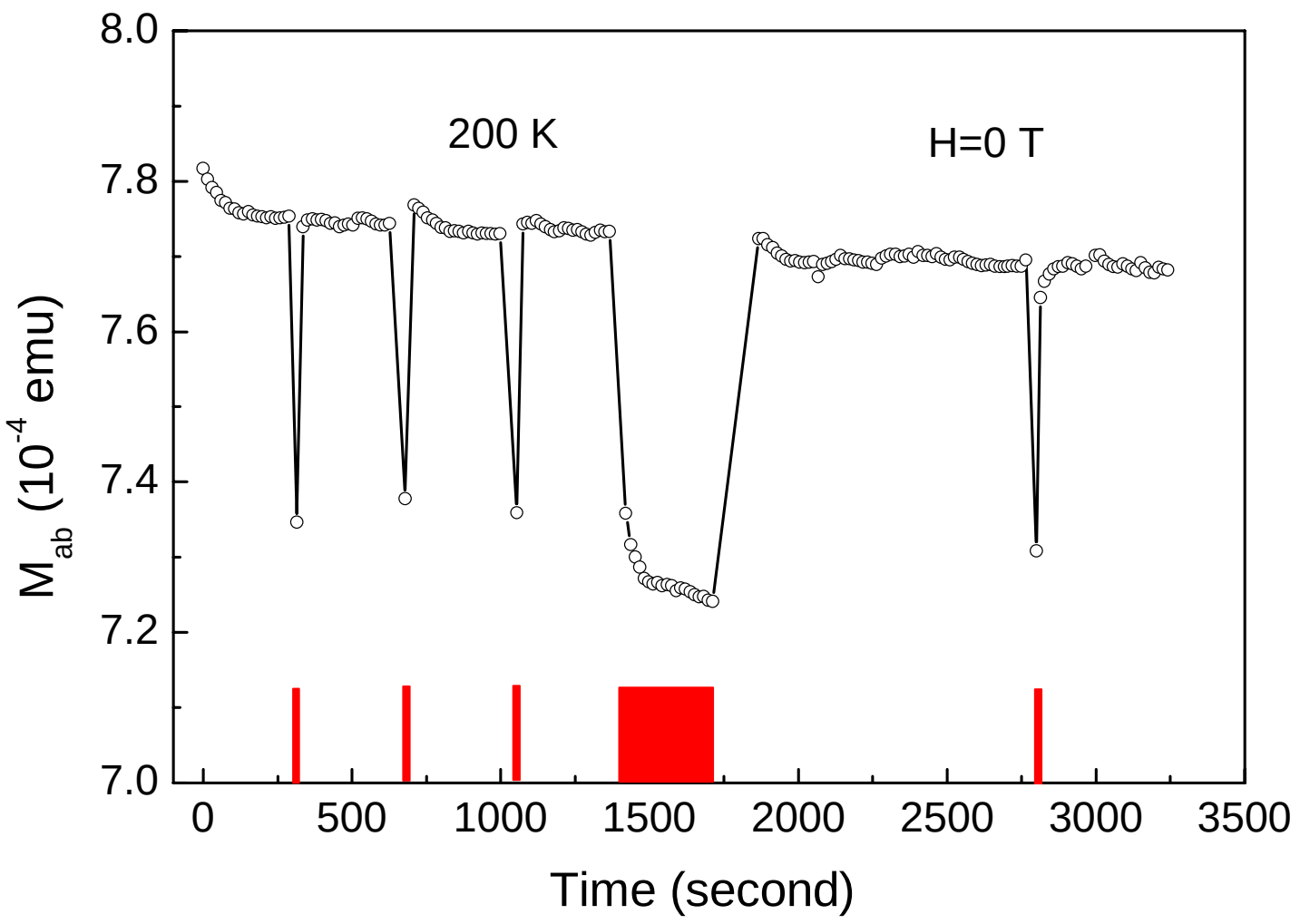